\documentclass[preprint]{elsarticle}

\usepackage{lineno,hyperref}
\usepackage{physics, amssymb}

\journal{Journal of Alloys and Compounds}

\begin{document}

\begin{frontmatter}

\title{Crystal structure and local ordering in epitaxial  Fe$_{100-x}$Ga$_x$/MgO(001) films}

\author[icma,dep]{Miguel Ciria \fnref{myfootnote1}}
\author[icma,dep]{Maria Grazia Proietti}
\author[icma,dep]{Edna C. Corredor }
\author[icma,dep]{David Coffey \fnref{myfootnote}} 
\author[icma,dep]{Adri\'an Begu\'e} 
\author[icma,dep]{C\'esar de la Fuente}
\author[ina,lma,dep]{Jos\'e I. Arnaudas}
\author[ina,lma]{A. Ibarra}
\address[icma]{Instituto de Ciencia de Materiales de Arag\'on, Consejo Superior de Investigaciones Cient\'{\i}ficas, Zaragoza, Spain.}
\address[dep]{Departamento de F\'{\i}sica de la Materia Condensada, Universidad de Zaragoza, Zaragoza, Spain.}
\address[ina]{Instituto de Nanociencia de Arag\'on, Universidad de Zaragoza, Zaragoza, Spain.}
\address[lma]{Laboratorio de Microscopias Avanzadas (LMA), Universidad de Zaragoza, Zaragoza, Spain}

\fntext[myfootnote1]{Corresponding author: miguel.ciria@csic.es}
\fntext[myfootnote]{{Current address: Delft University of Technology, 
Kavli Institute of Nanoscience,
Department of Quantum Nanoscience.}}




\begin{abstract}

In this work we present a study of the structural properties of Fe$_{100-x}$ Ga$_x$ (x$<$30) films grown by Molecular Beam Epitaxy on Mg0(100). We combine long range and local/chemically selective  X-ray probes (X-ray Diffraction and X-ray absorption spectroscopy)  together with real space imaging by means of Transmission Electron Microscopy and surface sensitive  \textit{in situ}  Reflected High Energy Electron Diffraction.  
For substrate temperature $T_s$ below 400 $^o$C we obtain \textit{bcc} films while, for \textit{x} $\approx$ 24 and $T_s \geq$ 400$^o$C the nucleation of the \textit{fcc} phase is observed. For both systems a Ga  anticlustering  or local range ordering phenomenon appears. The Ga/Fe composition in the first and second coordination shells of the \textit{bcc} films is different from that expected for a random Ga distribution and is close to a D0$_3$ phase, leading to a minimization of the number Ga-Ga pairs. On the other side, a long-range D0$_3$ phase is not observed indicating that atomic ordering only occurs at a local scale. Overall, the epitaxial growth procedure presented in this work, first, avoids  the formation of a long range ordered D0$_3$ phase, which is known to be detrimental for magnetostrictive properties, and second, demonstrates the possibility of growing  \textit{fcc} films  at temperatures much smaller than those required to obtain bulk \textit{fcc} samples.

\end{abstract}


\end{frontmatter}


\section{Introduction}

Iron-gallium alloys (Fe$_{100-x}$Ga$_x$) have become an important material for magnetostrictive applications because of their large tetragonal magnetostriction $\lambda_{100}$ at low field  for alloys around  18.4  \% Ga (Galfenol composition)\cite{Clark2000,Clark2001,Clark2003}, whereas,  at the same time,  they provide good corrosion resistance and mechanical hardness \cite{Atulasimha2011}. The interest in this alloy has been enhanced because the report of a non-joulian magnetostrictive behavior \cite{NatureFeGa} calls for further experimental and theoretical examination. Large  magnetoelastic (ME) coupling is a rewarding property in thin films and patterned elements, as the ME anisotropy can control the orientation of the magnetization \textbf{M}. Epitaxial Fe$_{100-x}$Ga$_x$ thin films have shown a remarkable potential for microwave \cite{Parkes2013} and energy conversion \cite{Onuta2011} applications, by using a piezoelectric layer to control the magnetic anisotropy through the modification of the strain in the magnetic layer by means of an applied voltage. This method, which has the advantages of low power consumption and efficiency in the \textbf{M} switching, has been proposed to  control  \textbf{M} in the magnetostrictive layer of MRAM devices \cite{Hu2011}. 
Improved magnetic properties  are obtained by the control of the crystalline phase as it has been pointed out by studies on rare-earth iron Laves phases alloys, showing that the magnetostriction can be enhanced in the phase boundary separating two ferromagnetic phases of different crystallographic symmetries \cite{PhysRevLett.104.197201, PhysRevLett.111.017203}. For Fe-Ga alloys,  the synergistic use of cubic phases with \textit{bcc} and \textit{fcc} symmetries conducts to  composites with stable magnetization and magetostriction at high temperature \cite{Ma2017}.

\begin{figure}
	\includegraphics[width=0.9\textwidth]{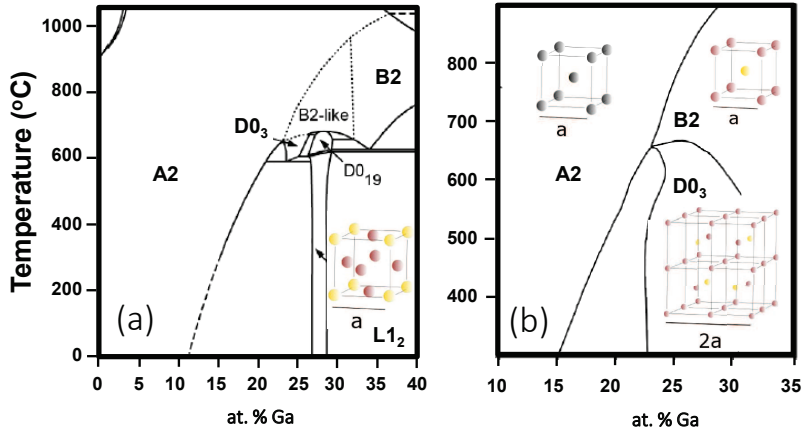} 
	\caption{(a) Equilibrium and (b) metastable phase diagram for Fe-Ga alloys adapted, respectively,  from references \cite{inCollFeGa} and \cite{IKEDA2002198}. The crystal structures for A2, B2, D0$_3$ and Ll$_2$ phases are also shown, with gray spheres indicating indistinctly Fe or Ga atoms while red and yellow colors specify, respectively, the positions of iron and gallium atoms (lines are guides for the eye).} 
	\label{fig:FiguraDiagramas}
\end{figure}

Bulk samples are obtained after processing the melted alloy by different routes that include or combine slow cooling, quenching and annealing, for a review see \cite{Handbook} and references therein. Regarding the stability of \textit{bcc} and \textit{fcc} crystal phases (see diagram in Fig. \ref{fig:FiguraDiagramas}a \cite{inCollFeGa}) it is established that the formation of the \textit{fcc} Ll$_2$ phase requires  a very well controlled procedure that includes long annealing times \cite{Srisukhumbowornchai2002}. Thus, the D0$_3$ phase can be present  at low temperature instead of a mix of \textit{bcc} and \textit{fcc} ordered phases as is observed in the metastable phase diagram of Fig. \ref{fig:FiguraDiagramas}b \cite{IKEDA2002198}. 

Thin film technology offers a different route to obtain materials and adds a new variable: Strain due to a substrate with a lattice parameter different from that of the film  can induce the nucleation of a phase  at lower temperature, as is observed for the canonical  Fe/Cu(100) system \cite{Liu1}. 
Another factor is the local diffusion of the species forming the alloy and its capacity to generate an ordered phase or remain at the sticking point with a disordered structure, either chemical or structural, during the process of adsorption of atoms at the films surface. \textit{bcc} Fe-Ga films can grow epitaxially on MgO(001) \cite{McClure2009}, a material used  in important  systems as tunnel junctions, on ZnSe(001) where a metastable next-neighbor pairing between Ga-Ga atoms has been reported \cite{Eddrief2011} and on GaAs(001) \cite{Parkes2013,Beardsley2017}. Here, we present a study of  Fe$_{100-x}$Ga$_x$ films grown on MgO(001)  crystals  with  x $<$ 30 and substrate temperature $T_s$ between 150 $^o$ C and 700 $^o$ C for a  growing rate of about 0.7 nm/min. Several characterization techniques are used to look into the long and short range structure of these epitaxial films: X-ray diffraction, RHEED, TEM and EXAFS. The main result is the observation of the nucleation of a \textit{fcc}  Ll$_2$ phase for \textit{x} $\approx$ 25 for  $T_{s}\gtrsim$ 400$^o$ C and the formation of metastable \textit{bcc} phases at $T_{s} \approx$ 150$^o$ C  for all the range of composition studied with an anticlustering mechanism of the Ga atoms.

\section{Experimental methods}

Fe$_{100-x}$Ga$_x$(001) epitaxial films were grown on MgO(001) by Molecular Beam Epitaxy. Tetragonal deformation is expected to occur due to the lattice mismatch with the MgO substrate. Bulk Fe$_{80}$Ga$_{20}$ and MgO have a lattice parameter a = 2.90 \AA\enspace and a =  4.21 \AA\enspace respectively. When FeGa is grown onto MgO (001), its lattice is turned by 45$^o$ in the growth plane in such a way that the FeGa [110] and MgO [100] directions are aligned. This provides a matching condition with the substrate that gives place to a tensile in-plane mismatch strain equal to 2.6\%. Prior to  growth,  the MgO(001) crystal was held at 800 $^o$C  to obtain a clean surface as shown by the Reflected High Energy Electron Diffraction (RHEED) patterns displayed in Fig. \ref{fig:FiguraRHEED}a-b.  These images display sharp spots and Kikuchi lines along the [100],  Fig. \ref{fig:FiguraRHEED}a, and [110], Fig. \ref{fig:FiguraRHEED}b, azimuthal directions. The deposition of Fe and Ga atoms was carried out by using an e-beam gun and a high temperature cell, respectively, with $T_s$ ranging between 150 $^o$C and 700 $^o$C.  The growth rate  was about  0.7 nm/min.  The film composition was obtained by means of dispersive X-ray spectroscopy (EDX), and the thickness was determined by X-ray reflectivity. A 2 nm thick Mo  capping layer was deposited onto the Fe-Ga layers to prevent from  oxidation. The lattice parameters were determined by means of XRD measurements using a Bruker D8 Advance High resolution diffractometer  and a Rigaku rotating anode D/max 2500 diffractometer working with a Bragg-Brentano configuration. Transmission electron microscopy images were obtained using a Tecnai F30 Microscope in a sample thinned down by a Helios 600 dual beam system.

The EXAFS spectra  were recorded at the Fe K-edge (7112 eV), and at the Ga K-edge (10367 eV) at room temperature, in fluorescence mode.
The  experiments were performed on beamline BM30B at the European Synchrotron Radiation Facility (ESRF), Grenoble, France  and on  beamline XAFS at the Italian Synchrotron Radiation Facility  (Elettra), Trieste, Italy.  Measurements were performed with 
the samples  surface  oriented nearly parallel to the incident beam $\vec{\varepsilon}_{\|}$ (incidence angle equal to about $5^{\circ}$)  and perpendicular to it $\vec{\varepsilon}_{\bot}$ (incidence angle equal  to about $85^{\circ}$).
A pure \textit{bcc} Fe film grown epitaxially on MgO was also measured as reference. 

The films studied  are listed in Table \ref{tab_estruc}. The first column is a code that, hereafter, is used to refer to the samples studied in the following sections. The samples are divided in two sets: S1 and S2. S1 corresponds to films grown at $T_{s}$ = 150 $^o$C and \textit{x} varying between 0 and 28, a value that is used in the label. For the second set of films, S2, the substrate temperature varies form 300 $^o$C to 700 $^o$C, being used as a label, and the flux of Fe and Ga beams is fixed  except for the sample with \textit{x} = 13  and  $T_{s} $ = 600 $^o$C, which was grown to look for the presence of \textit{fcc} phase at low gallium content.

\begin{table}
	\begin{tabular}{cc|cccccc}
		
		&       &  &   &  &   &    &\\
		\textit{Sample} &x  & SL peak & [100] & [110] & a$_{\bot}$ & a$_{\|}$   & th      \\
		 & (at \% Ga) &  &  &  &  (\AA)& (\AA)   &  (nm)    \\
	
		\hline 
		&       &  &   &  &   &    &\\
		S1-0 &  0   & N   & N  & N & 2.886  &         &44   \\ 
		S1-12  & 12    & N & N  & N & 2.866  &         &25   \\
		S1-13  & 13  & N   & N  & N & 2.871  &         &     \\
		S1-18  & 18   & N  & N  & N & 2.879  & 2.94    &14.5 \\
		S1-21   & 21  & Y  & N  & N & 2.890  & 2.930   &16   \\
		S1-22   & 22 & Y  & N  & Y & 2.884  & 2.947   &17   \\
		S1-24a   & 24 & Y   & Y  & Y & 2.888  & 2.938   &20   \\
		S1-24b  & 24  & Y   & Y  & Y & 2.903  &         &24.3 \\
		S1-28a  & 28  & Y   & N  & Y & 2.894  &         &20.5 \\
		S1-28b  & 28  & Y   & Y  & Y & 2.904  & 2.918   &27.6 \\
		&       &  &   &  &   &    &\\	
		\hline\hline
		&       &  &   &  &   &    &\\
		\textit{Sample}    & $T_{s}$& x & str.& SL peak  & a$_{\bot}$  & a$_{\|}$   & th    \\
			   & ($^o$C)&  (at \% Ga)& &   & (\AA) &  (\AA)  & (nm)    \\

		\hline
		&& & &      &  &    & \\
		S2-300    &300& 24$^+$     &bcc   &Y  &  2.935  &         &24    \\
		S2-400    &400& 24 &fcc   &N  &  3.614  &         &27.1  \\
		S2-500    &500& 24 &bcc   &Y  &  2.909  & 2.898   &26.5  \\
		S2-600    &600& 24&fcc   &Y  &  3.687  & 3.677   &-     \\
		S2-600b   &600& 13&bcc   &N  &  2.900  &         &      \\
		S2-700    &700& 24&bcc   &Y  &  2.907  &         &      \\
		\hline
		
	\end{tabular}
	\caption {List of the films studied in this work. Films with the S1 label were grown at $T_{s}$ = 150 $^o$C  and have \textit{bcc} structure. From left to right: Sample code, gallium content (EDX), presence of the (001) SL peak,  reconstruction along the FeGa [100] and [110] azimuthal directions (RHEED), out-of-plane and in-plane lattice parameter (XRD) and film thickness th (XRR). Films labeled with S2, except S2-600b,  were prepared with the same experimental conditions for the evaporation beams and  the composition of the alloy is assigned to the value obtained by EDX for film S2-300 (+). Sample S2-600b was prepared with a lower Ga/Fe crucible flux ratio to obtain a lower Ga content. From left to right: sample code, gallium content (EDX), crystal structure,  presence of the (001) SL peak, out-of-plane and in-plane lattice parameter (XRD) and film thickness th (XRR). Errors on the EDX data are about $\pm$ 1\%.} 
	\label{tab_estruc}
\end{table}

\section{Experimental Results}

\subsection{Reflected high energy electron diffraction} 
The \textit{in situ} RHEED technique provides an initial evaluation of the growth  mechanism and the layers crystal structure. The  RHEED patterns for representative films with \textit{bcc}  and \textit{fcc} structure taken with the incident e-beam along the MgO [100] and [110] directions, see Fig. \ref{fig:FiguraRHEED}, show a $\pi/2$ periodicity indicating that the Fe-Ga films have a fourfold symmetry. 
The images obtained for the film S1-13, see  Fig.\ref{fig:FiguraRHEED}c-d, are identical to those obtained for reference Fe films (not shown), suggesting a \textit{bcc} structure. The distance ratio between the main reflections of these films along the  MgO [100] and [110] directions is  about 1.4.  These lines, that correspond to atomic  planes distances, are  aligned with the reflections coming from the  MgO [100] and [110] directions. RHEED images taken on films  S1-24 and S1-28a show  extra lines  indicating surface reconstructions that will be discussed in the next section. 
On the contrary, for the film S2-600, Fig. \ref{fig:FiguraRHEED}i-j,  the  distance ratio between the main reflections along the  MgO [100] and [110] directions  is 0.7 ($\approx 1/1.4 $)
and  the stronger streaks in Fig. \ref{fig:FiguraRHEED}i-j, are no longer aligned with the  substrate lines. These significant differences in the RHEED patterns suggest that the crystal structure of the S2-600 film is markedly different from the other Fe-Ga and pure Fe films, although it has a fourfold symmetry plane on the MgO(001) surface. The latter structures correspond to a \textit{fcc} phase as determined by XRD experiments reported in the following section. The epitaxial relationships are  MgO[100]$\|$FeGa[110] for \textit{bcc} films and MgO[100]$\|$FeGa[100] for \textit{fcc} films.

\subsubsection{RHEED Superstructures}

The RHEED images taken on the \textit{bcc} films grown at $T_s$ = 150 $^o$C  show that by increasing the Ga concentration, a complex superstructure is observed along the two main in-plane directions. Films grown at higher temperature can display  other reconstructions that are not discussed in this work, because of the complexity of the subject. For instance  the c(2 $\times$ 2) reconstruction,  that for Fe-Ga alloy could be interpreted as ordering of the Ga and Fe atoms, is also observed in reference pure iron films grown at T$_{s}>$ 220 $^o$C \cite{Oka2001}. 

\begin{figure}
	\includegraphics[width=0.9\textwidth]{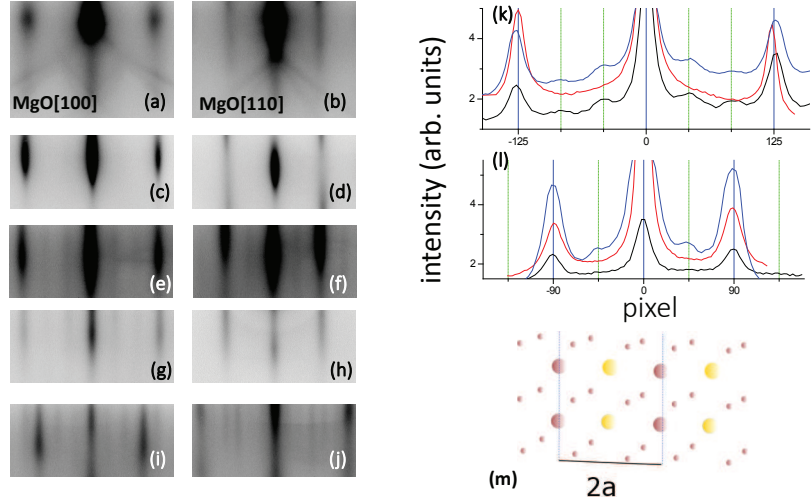}
	\caption{RHEED patterns of the clean MgO(001) substrate  along the  MgO $<$100$>$ (a)  and $<$110$>$ (b) and the corresponding images for \textit{bcc} films  (c-d) S1-13;(e-f) S1-24; (g-h) S1-28a; and (i-j) the \textit{fcc} film S2-600 (x=24). Intensity \textit{vs.} pixel position for the \textit{bcc} images along the MgO[100](k) and MgO[110] (l) directions for films S1-13 (red line), S1-28a (black line) and S1-24 (blue line). The intensity of the lines has been shifted vertically for clarity. (m) The large round objects represent a sketch of a two dimensions area that could explain a (2 $\times$ 1) reconstruction:   Alternate rows of Ga and Fe atoms. The color code is the same as shown in Fig. 1. and the smaller balls are added to complete a \textit{bcc} structure.} 
	\label{fig:FiguraRHEED}
\end{figure}

The RHEED images for  film S1-13 do not show the presence of  surface reconstructions, however for film  S1-24, 
along the FeGa $<$110$>$ azimuth  [Fig. \ref{fig:FiguraRHEED}e]  two weak lines can be observed  between the main reflections, whereas  for the image taken along the FeGa $<$100$>$ azimuth [Fig. \ref{fig:FiguraRHEED}f]  a line can be observed at  halfway between the stronger strikes.  Film  S1-28a the reconstruction lines are absent in the image taken along FeGa $<$100$>$ azimuth [Fig. \ref{fig:FiguraRHEED}h] but not along FeGa $<$110$>$ azimuth [Fig. \ref{fig:FiguraRHEED}g].  This fact is highlighted in the profiles presented in Fig. \ref{fig:FiguraRHEED}k and l showing the intensity \textit{vs.} pixel position perpendicularly to the strikes of the images in Fig. \ref{fig:FiguraRHEED}c-h taken in films with and without reconstruction  along both azimuthal directions. 

The intensity of the SL lines is stronger for  images taken along the [110] directions that for images obtained along the [100] directions, which implies that for some films the latter is undetected or with small  intensity value.  This fact suggests  that the presence of lines  along the [100] and [110] directions may obey to different reasons. 

The super-order along the [110] direction can be interpreted  as due to a set of domains  with basis vector (0,3) and (1,-1) with respect to square unit cell. This diffraction pattern  has  similarities with those  shown in shape memory alloys, where the cubic lattice undergoes tetragonal transformation and structural domains share the [110] direction from the austenite phase in the martensite  phase of the NiMnGa alloy resulting in superlattice peaks  along the austenite [110] direction \cite{Luo2011}.  

The weak lines observed along [001] directions without spot at the middle distance of the main [110] reflections  can indicate a reconstruction (2 $\times$ 1) with domains doubling the lattice parameter \textit{a} along both in-plane [100] and [010] directions. These reflections are very weak and wide, suggesting a dilute presence of these objects with small lateral size. We speculate with the fact that the residual small amount of Ga-Ga nearest neighbors could generate 
a 2\textit{a}  periodicity in real space (see Fig. \ref{fig:FiguraRHEED}.m) giving rise to the diffraction spots observed in Fig. \ref{fig:FiguraRHEED}. The fact that these spots are not observed for films with low content of Ga  may indicates that the number of Ga dimers must be low or/and disordered in the Fe matrix; on the other hand, it appears that films with large content of Ga (S1-28a) could also shown an absence of the (2 $\times$ 1) reconstruction suggesting that increasing the concentration of  Ga is  detrimental for the formation of Ga-Ga nearest neighbors as happens  for the D0$_3$ structure.

\begin{figure}
	\centering
	\includegraphics[width=0.9\linewidth]{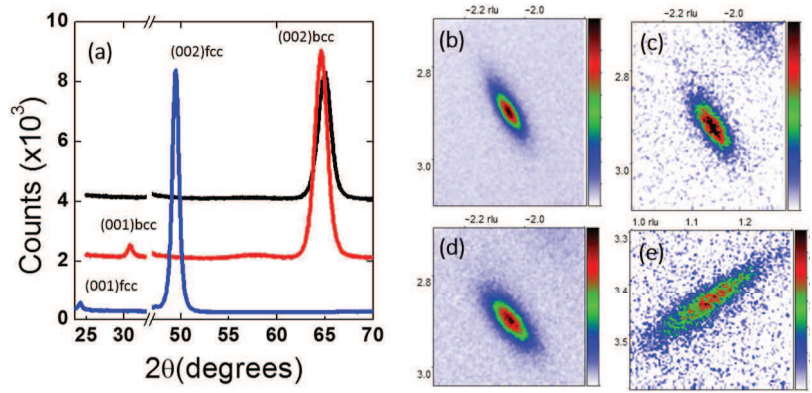}
	\caption{(a) XRD $\theta/2\theta$ scans  for  films S1-13 (black line), S1-24a (red line) and S2-700 (blue line). Reciprocal space map for films (b) S2-500 and (c) S1-24  (d) S1-28b taken around the($\overline{22}4$) reflection and (e) film S2-600 at the ($\overline{1}\overline{1}3$) one.  Units in coordinate axes of the maps are multiples of  2$\pi$/a$_{MgO}$, with a$_{MgO}$ the MgO lattice parameter. The number of counts (arbitrary units) is associated with a color code located at the right side.}
	\label{fig:FiguraStruc_1}
\end{figure} 

\subsection{X-ray diffraction}

The samples have been studied by performing X-ray diffraction with the scattering vector normal to the film  surface. Representative X-ray diagrams are presented in Fig. \ref{fig:FiguraStruc_1}a, showing  the main reflection along the growth direction  and in some cases  superlattice peaks due to chemical order (labeled as (001) in Fig. \ref{fig:FiguraStruc_1}a). 

The \textit{bcc} structure is assigned to  films with  main reflections located at 2$\theta\approx$ 65$^o$, which correspond  to the  \textit{bcc} (002) peak.  The \textit{fcc} structures is associated to  films with a main  reflection at 2$\theta\approx$ 49$^o$ corresponding the the \textit{fcc} (002) peak. For the \textit{bcc} films the superlattice peak is located at 2$\theta\approx$ 31$^o$ while for the \textit{fcc} films that peak appears at  2$\theta\approx$ 24.5$^o$. No evidence of other reflections is found in these scans. The lattice parameters obtained for both structures (see Table \ref{tab_estruc}) are within the range of values observed in the bulk  alloys: 2.90 \AA\  and 3.68 \AA\ for \textit{bcc} and \textit{fcc} crystals, respectively\cite{Kawamiya}.

The A2 structure, with Fe and Ga atoms randomly distributed in the lattice positions does not present the (001) superlattice peak. 
Two \textit{bcc} phases, B2 structure  and the D0$_3$ (including a modified D0$_3$\cite{He2016177}), in which the Ga atoms are ordered with respect to the Fe matrix, could  explain the presence of the (001) reflection. XRD scans were performed to observe superlattice peaks  for the D0$_3$ \textit{bcc} symmetry as the (1/2 1/2 1/2) reflection but no signal was detected. Nevertheless,  the presence of the (001) clearly indicates a chemical order. For the \textit{fcc} films, the equilibrium phase diagram, see Fig. \ref{fig:FiguraDiagramas}a, shows that a Ll$_2$ structure exists in a certain range of temperature and composition. The presence of this ordered  phase implies that the (001)-\textit{fcc} reflection is no longer extincted, as observed for sample S2-600 that displays  a weak  peak at 2$\theta\approx$ 24.5$^o$.

For the S1 samples ($T_s$ = 150$^o$ C)  the presence of the (001) SL peak and  the  surface reconstruction seem to be associated with the increment of the Ga content, see Table \ref{tab_estruc}. Sample S1-21 is an exception to this behavior, an issue that can be due to the fact that the intensity of the (001) peak, compared with the other samples, takes its lowest value  and the RHEED reconstruction signal could be overlooked because it is below the sensibility achieved during that \textit{in situ}  experiment.

Asymmetrical scans, see Fig. \ref{fig:FiguraStruc_1}b-e, confirm the epitaxial relations observed by RHEED and allow determining the average strain in the film plane. Fig. \ref{fig:FiguraStruc_1} shows $<$224$>$ reflections for \textit{bcc} and $<$113$>$ reflection of the  \textit{fcc}  film.  These data  indicate that the \textit{bcc} films undergo an in-plane extension and out-of-plane compression to accommodate their lattice parameter with the MgO substrate.  We observe that built-in strain can be relaxed both by increasing $T_s$ and film thickness.

\subsection{Crystal phase diagram}
\begin{figure}
	\includegraphics[width=0.7\textwidth]{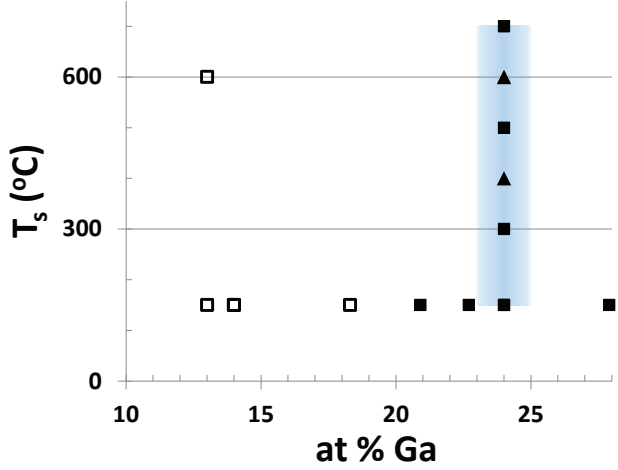}
	\caption{Phase diagram obtained from the analysis of the structural measurements. Triangles stand for \textit{fcc} phases, empty squares for A2 \textit{bcc} phase and solid squares for \textit{bcc} with superlattice peaks. The shaded area shows the error in composition ($\Delta$x$\cong$ 1).}
		\label{fig:FiguraDia}
	\end{figure}

Diffraction experiments reveal in some  samples the presence of weak (100)  superlattice reflections that indicate a chemical  ordering of  Ga in the Fe lattice. Combining XRD and RHEED measurements, it can be concluded that films with  patterns like that shown in Fig. \ref{fig:FiguraRHEED}c-h have a \textit{bcc} structure, while films that present patterns like that shown in Fig. \ref{fig:FiguraRHEED}i-j are cubic with a \textit{fcc} structure.
Different symbols are also used for  \textit{bcc} films  if the (001) superlattice peak is observed by the XRD measurement. 
Considering these results, a phase diagram is presented in Fig. \ref{fig:FiguraDia}, where the symbols indicate the phase observed at a certain value of film composition and substrate temperature $T_s$. It is also noteworthy that for the films grown with the same nominal composition ($x=24$) both the  \textit{fcc} and \textit{bcc} structures are  observed, suggesting that a  minute fluctuations in the composition can induce the nucleation in the whole film of a single cubic phase. 

\subsection{Transmission Electron Microscopy}
\begin{figure}
	\centering
	\includegraphics[width=0.8\linewidth]{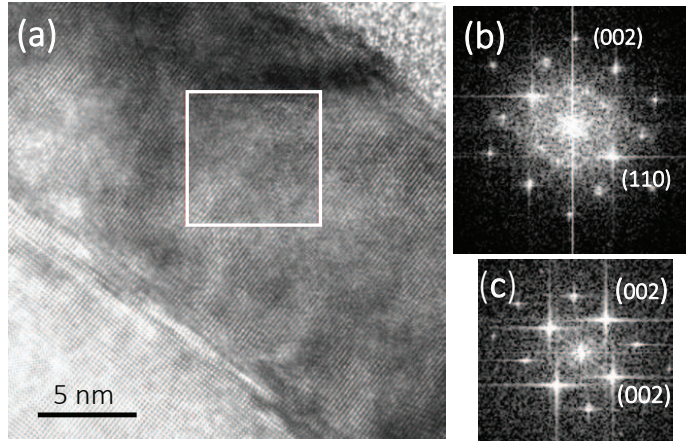}
	\caption{(a) Transmission electron microscopy image of the   S1-24 
		film showing the MgO and FeGa layer perpendicular to the [110] direction. (b) FFT of the area marked in (a), corresponding to the FeGa film and (c) FFT of the MgO.}
	\label{fig:FiguraStruc_2}
\end{figure}

Transmission electron microscopy image of the S1-24 film is shown in Fig. \ref{fig:FiguraStruc_2}a with the MgO(100) and FeGa(110) planes normal to the electron beam. The image displays  rows of atoms that demonstrate  the crystalline character of the film and the epitaxial relationship with the substrate. 

The fast Fourier transformation of the Fe-Ga  marked area of the TEM image displayed in Fig. \ref{fig:FiguraStruc_2}a is presented in Fig. \ref{fig:FiguraStruc_2}b and corresponds to the [110] zone. The spots with the (002) and (110) labels mark the growing and the in-plane direction, respectively. At  the position corresponding to the (001) reflection a double peak is clearly present. Regarding the (1/2 1/2 n/2) reflections, the signal is absent for any value of \textit{n}, either odd or even. This analysis has been performed on different areas of the film and most of the them present the same structure although the splitting of the (001) varies slightly from point to point. The splitting of diffraction spots in alloys has been explained as a result of the presence of anti-phase boundaries \cite{DoublePeak}
that split the reflections due to a superlattice structure but not those due to the basic lattice. The lack of the (1/2 1/2 n/2) reflections suggest that the ordered FeGa phase does not correspond to a D0$_3$ or the modified  D0$_3$ structures \cite{He2016177}, corroborating the X-ray diffraction data. 

\subsection{Extended X-ray Absorption Spectroscopy.}

Extended X-ray Absorption Spectroscopy (EXAFS) has been used to determine the local atomic environment of Ga and Fe by quantitative analysis of the oscillatory contribution to the X-ray absorption spectrum showing up above the X-ray absorption edge of Ga and Fe. 
The interested reader can find a recent review on the fundamentals of this technique in reference \cite{Rehr} and its application to Materials Science in reference \cite{Boscherini-Chap7-Lamberti-Agostini-2012}  and references therein. EXAFS can provide interatomic distances, thermic/static disorder factors (Debye-Waller) and chemical nature of Nearest Neighbors (NN) and Next Nearest Neighbors (NNN) giving the sample composition at local scale. This allows one to discern if the Ga atoms are randomly distributed in the lattice or some atomic ordering mechanism occurs. In addition, we performed EXAFS experiments with
the beam polarization vector, $\vec{\varepsilon}$, directed along the $[100]$ and $[001]$ crystallographic directions, corresponding to $\vec{\varepsilon}_{\|}$ and  $\vec{\varepsilon}_{\bot}$ respectively to exploit the strong anisotropy of the EXAFS spectrum \cite{brouder_ang-dep_1990}  and  determine the in-plane and out-of-plane lattice parameters, the distribution of Ga-Ga pairs and possible anisotropies in the local Ga concentration.  

\subsubsection{EXAFS analysis}

We performed a fit procedure of the experimental EXAFS spectra to theoretical EXAFS signals calculated by using \textit{ab-initio} theoretical phases and amplitudes. 
To this end we generated four clusters composed by 113 and 78 atoms for the \textit{bcc} and \textit{fcc} symmetry respectively, with a radius of $6.1$~\AA\ by the TKATOMS code \cite{Ravel2}. The lattice parameters were set equal to the pure \textit{bcc} and \textit{fcc} Fe structures, i.e. 2.870 \AA\  and 2.531 \AA\ respectively. The absorber central atom was either Fe or Ga and exclusively Fe or Ga as scatterer atoms.
Theoretical amplitudes and phase shifts were calculated  \textit{ab-initio} by \texttt{FEFF8} code \cite{ankudinov_real-space_1998} for the model clusters taking into account $\vec{\varepsilon}$.
The theoretical EXAFS signals for Fe-Ga (Fe K-edge) or Ga-Fe (Ga K-edge) systems were obtained by combining theoretical phases and amplitudes of the pure Fe and pure Ga clusters with a population factor \textit{x}. 

Atomic background subtraction in the EXAFS region was performed by  \texttt{AUTOBK} code implemented by the \texttt{ATHENA} graphical interface \cite{ravel_athena_2005}. Fit of theoretical signal to EXAFS was performed by using the \texttt{IFEFFIT} \cite{newville_ifeffit_2001} code implemented by the \texttt{ARTEMIS} interface \cite{ravel_athena_2005}.

\begin{figure}
	\includegraphics[width=0.9\textwidth]{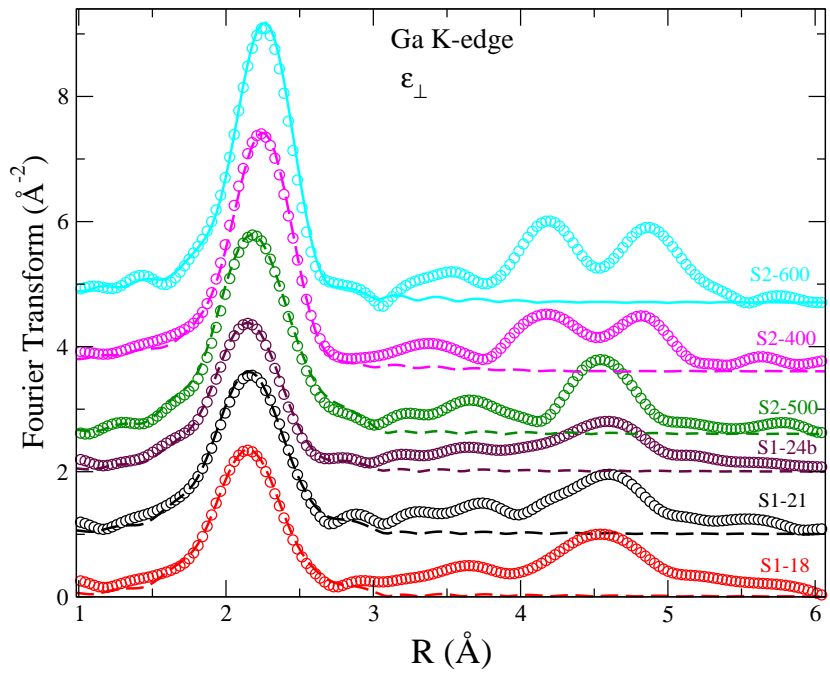}
	\caption{ (Color online) FT amplitude of $k^2\chi(k)$  EXAFS signals (circles) with out-of-plane ([001])
		X-ray beam polarizations, at Ga $K$ edge,  for several Fe-Ga alloys together with the correspondent best fit curves (solid lines) performed in \textit{q} space. Curves are vertically shifted for clarity.}
	\label{EXAFS_Ga}
\end{figure}

\begin{figure}	
	\includegraphics[width=0.8\textwidth]{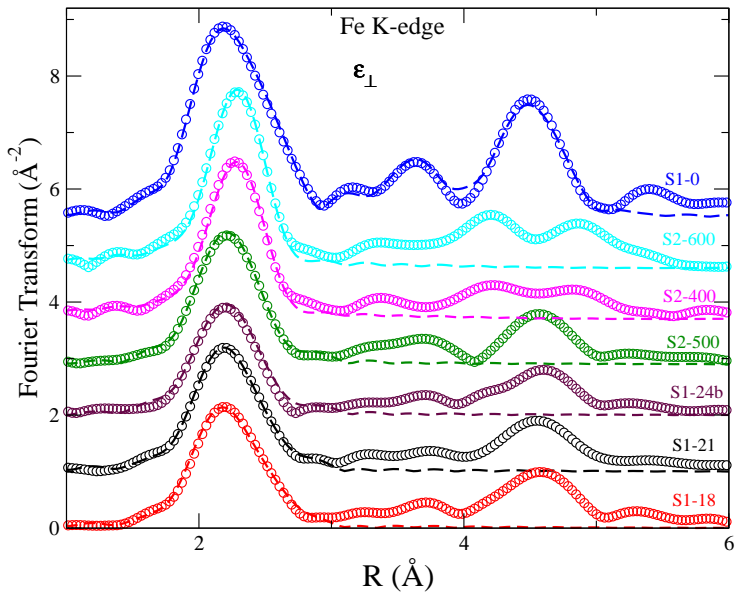}
	\caption{ (Color online)  FT amplitude of $k^2\chi(k)$  EXAFS signals (circles) with out-of-plane ([001])
		X-ray beam polarizations, at Fe $K$ edge, for several Fe-Ga alloys together with the correspondent best fit curves (solid lines) in \textit{q} space. Curves are vertically shifted for clarity.}
	\label{EXAFS_Fe}
\end{figure}

The Fourier Transformed (FT) EXAFS spectra are shown in Fig.  \ref{EXAFS_Ga} (Ga K-edge) and Fig. \ref{EXAFS_Fe} (Fe K-edge) for some of the samples of Table \ref{tab_estruc}. An EXAFS spectrum of a pure \textit{bcc} Fe film grown epitaxially on MgO is also reported for comparison in Fig. \ref{EXAFS_Fe}.
In this study we restrict the fit procedure to the first peak of the FT spectrum showing up in the \textit{R} region from  $1$~\AA\ to  $3$~\AA. It corresponds, for the \textit{bcc} structure, to the contribution of the I and II coordination shells, i.e. to the atoms at center and at the corner of the \textit{bcc} cube respectively, that due both to the small difference in distance from the central atom and to the limited \textit{R}-space resolution associated to the limited spectrum \textit{k}-range, cannot be resolved. For the \textit{fcc} structure, it corresponds to a single shell of 12 NN situated at the center of the \textit{fcc} cube faces.
The FT have been calculated in the range $2-12.5$~\AA$^{-1}$ and provide by sight qualitative information on the Fe and Ga short range order environment. For most of samples the FT spectra are similar to that of pure \textit{bcc} Fe showing that the \textit{bcc} symmetry is maintained.
Nevertheless, when the substrate temperature reaches 400 $^o$C, the \textit{fcc} symmetry becomes possible and a phase change can take place as also observed by diffraction and reported in the previous section: Samples S2-400 and S2-600 have a \textit{fcc} phase.

\subsubsection{bcc samples}

The fitting was performed on the Fourier filtered $\chi(q)$ in the range $2-12.5 $~\AA$^{-1}$ refining
the following parameters: the origin of photoelectron energy $E_{0}$; the interatomic distances
$d_{I(Fe-Fe)}$, $d_{I(Fe-Ga)}$ (Fe K-edge), $d_{I(Ga-Ga)}$, $d_{I(Ga-Fe)}$ (Ga K-edge), $d_{II\bot(Fe-Fe)}$, $d_{II\|(Fe-Fe)}$, $d_{II\bot(Fe-Ga)}$, $d_{II\|(Fe-Ga)}$ (Fe K-edge), $d_{II\bot(Ga-Ga)}$, $d_{II\|(Ga-Ga)}$,  $d_{II\bot(Ga-Fe)}$, $d_{II\|(Ga-Fe)}$ (Ga K-edge); the Debye-Waller factors for I and II shells, $\sigma_{I}^2$  and $\sigma_{II}^2$; the Ga population factor in the I and II coordination shell \textit{y} and \textit{z}. The $E_{0}$ best fit values for the different samples were very close to each other ($E_{0}\cong 1eV$), showing differences lower than 0.5 eV. The  $S_{0}^{2}$  amplitude reduction factor was fixed to 0.7 for all the samples; the coordination numbers were kept fixed to their crystallographic values for the \textit{bcc} system, \textit{i.e.}  $N_{I} = 8$ and $N_{II} = 6$. Note that the interatomic distances $d_{II\|(\bot)}$ correspond to the lattice parameters $a_{\|(\bot)}$ reported by XRD, but, due to the EXAFS chemical selectivity they are specific for Fe-Ga and Ga-Ga pairs and not average as for diffraction.   
The Ga concentration factor \textit{x} was split out into \textit{y} and \textit{z}, for the I and II shells respectively. If \textit{y} were found to be equal to \textit{z}, and equal to \textit{x}, \textit{i.e.} the nominal Fe$_{100-x}$Ga$_{x}$ composition,  the Ga distribution would be random; other values of \textit{y} and \textit{z} indicate ordering or clustering phenomena.

The possibility of non-random or ordered Ga distribution has to be taken into account according to previous EXAFS \cite{Pascarelli2008} and  XRD results \cite{Du2010}. We underline that in previous papers the FeGa samples studied were bulk samples obtained with different methods. In our case all the samples are thin epitaxial samples with very different growth dynamics compared with bulk samples and  the  phase diagram of which has never been studied.

\paragraph{Ga K-edge}

The fit results are reported in Table \ref{tab_Ga}.
The concentration of Ga atoms in the I and II coordination shells, \textit{y} and \textit{z}, around the Ga absorber, is found to be equal to 0 within the  fit absolute error bar ($\Delta{y} = \Delta{z} =$ 10) for all the samples except S1-24a, in which they both approach 10. No difference was  observed for \textit{y} and \textit{z}, regardless of the beam polarization. It has to be compared with the values found by EDX reported in Table \ref{tab_estruc} for \textit{x} that range from 12 to 28. This result shows a clear tendency of Ga to undergo an anticlustering/ordering mechanism since the Ga concentration at a local scale is remarkably lower than the average value obtained by EDX. Imposing the presence of one Ga atom in the I or II shell, as reported in ref \cite{Pascarelli2008}, always produced in our case an increase of \textit{R}-factor. 
This difference between the expected number of Ga NN atoms corresponding to a random distribution of Ga in the Fe lattice and the experimental value found by EXAFS can be associated to the presence of ordered B2 or  D0$_3$ phases at local scale  (B2-like and D0$_3$-like) for  which the  identity of NN and NNN correspond to the values expected for these phases. 
The values expected according to  D0$_3$ and B2 ordered clusters are \textit{y} = \textit{z} = 0 and \textit{y} = 0, \textit{z} = 100 respectively. It can be easily understood looking at Fig. \ref{fig:FiguraDiagramas} that shows the D0$_3$ crystal structure typical of Fe$_3$Al (space group Fm3m)  together with the B2 structure typical of AuCd (space group Pm3m).
In the D0$_3$ structure A2 (pure \textit{bcc} Fe) and B2 cells are stacked alternatively in each direction forming an \textit{fcc} structure with a doubled cell parameter. A full B2 ordering could take place with a Ga overall ratio of 50 at. \% Ga and a D0$_3$ structure could occur with 25  at. \% Ga, that is close to the Ga content of our samples. 
We can state then that the short range order Ga distribution tends for all the samples studied to a D0$_3$-like structure.
Concerning interatomic distances, the $d_{I(Ga-Fe)}$, values range from 2.50 \AA\ to 2.53~\AA, being close to the values found 
for bulk samples with about the same concentration.  $d_{II\bot(Ga-Fe)}$ and $d_{II\|(Ga-Fe)}$ range
from 2.84 \AA\ to 2.88 \AA\  and from 2.88 \AA\ to 2.94 \AA, respectively, showing that $d_{II\bot(Ga-Fe)} < d_{II\|(Ga-Fe)}$ in the samples grown with $T_{s}$ = 150 $^{o}$C. This indicates the presence of residual tensile in-plane strain due to the mismatch with the MgO substrate, that produces an out-of-plane compression and  tends to relax at higher $T_s$.

	\begin{table}
	\begin{tabular}{c|cccccc}
		\textit{Sample}   & $d_{I(Ga-Fe)}$  & $d_{II\bot(Ga-Fe)}$ & $d_{II\|(Ga-Fe)}$& \textit{y}=\textit{z} &$\sigma^2_{I}$&$\sigma^2_{II}$  \\
		
		&  (\AA)    &   (\AA) &  (\AA) &  (at \% Ga)   &  (\AA${^2}$)   & (\AA${^2}$)   \\
		\hline 
		S1-21      &2.521  &2.88  &2.92  &0 &0.008 & 0.02\\
		
		S1-18      &2.525  &2.85  &2.91  &0  & 0.008&0.017 \\
		
		S1-13     &2.53  &2.85  &2.90  &0    &0.007 & 0.014\\

		S1-24a    &2.512 & 2.87 & 2.94 & 10 & 0.008&0.025 \\
		
		S2-500   &2.53  &2.88  &2.88   &0   &0.006 &0.014\\
		
		S2-600    &2.53  &2.88  &2.88  &0  &0.006 &0.011\\
		\hline 
	\end{tabular}
	\caption {Best fit results at Ga K-edge. The perpendicular and parallel polarization spectra were fitted simultaneously. The interatomic distances refer to Ga-Fe pairs since no Ga-Ga pairs were observed for all samples except S1-24a in which the low Ga-Ga pairs presence does not allow one to determine the Ga-Ga distances that are set equal to Ga-Fe. Statistical error, calculated by the fit covariance matrix,  are equal to 0.01 \AA~ for $d_I$, 0.02 \AA~ for  $d_{II\bot/\|}$  and 0.005 for $\sigma^2$.}
	\label{tab_Ga}
\end{table}

\begin{table}
	\begin{tabular}{c|ccccc}
		\textit{Sample}    & $d_{I(Fe-Fe)}$       & $d_{II\bot(Fe-Fe)}$      & $d_{II\|(Fe-Fe)}$     &$\sigma^2_{I}$          &$\sigma^2_{II}$             \\
		&                  (\AA)  &                    (\AA) &                  (\AA)&                    (\AA$^2$) &               (\AA${^2}$)  \\
		\hline	
		S1-0  &2.467(4)    &2.854(7)  &2.854(8)      &  0.0045(5) & 0.05(1)\\
		S1-13     &2.46 &2.84  &2.88    &    0.005 & 0.014\\
		S1-18     &2.47  &2.85  &2.91   &  0.005&0.014 \\
		S1-21     &2.48  &2.87       &-        &   0.004 & 0.018\\		
		S1-24a     &2.49  & 2.87 & 2.92   &   0.008&0.019 \\
		S2-500      &2.47  &2.93 &2.91   &  0.006 &0.019\\
		S2-600     &2.471  &2.88  &2.88   &0.007 &0.019\\
		\hline 	
	\end{tabular}
	\caption {Best fit results at Fe K-edge. The perpendicular and parallel polarization spectra were fitted simultaneously. The interatomic  distances $d_{II\bot/\|}$ refer to Fe-Fe pairs.  Statistical error, calculated by the fit covariance matrix,  are equal to 0.01 \AA~ for $d_I$, 0.02 \AA~ for a $d_{II\bot/\|}$, 10 for \textit{z}  and 0.005 for $\sigma^2$.For sample S1-21  $d_{II\bot(Fe-Fe)}$ is not present because the specter with $\vec{\varepsilon}_{\|}$ could not be registered.}
	\label{tab_Fe}
\end{table}

\paragraph{Fe K-edge}

The analysis of the Fe  K-edge should confirm the findings  at the Ga K-edge, suggesting atomic ordering of Ga. According to a D0$_3$ ordered structure, the iron atoms have two different Wyckoff positions in the lattice, 8(c) and 4(b). For the 8(c)  site, there are 4 Ga atoms out of 8 in the I shell and 0 Ga atoms on the II shell, while  the atom at site 4(b) has 0 NN Ga atoms and 6 out of 6 as NNN. Therefore, the average values of \textit{y} and \textit{z}  would be equal to 33.

On the other side, the fit sensitivity to \textit{y} and \textit{z}  at the Fe K-edge is low, as also observed by other authors \cite{Pascarelli2008}
and is not possible to give a reliable value for these parameters.
As for the Ga K-edge, no difference was observed depending on polarization. 
The fit results are reported in Table \ref{tab_Fe}. The Fe-Fe interatomic distances have been refined keeping the correspondent Fe-Ga distances fixed to the values found at the Ga K-edge ($d_{I(Ga-Fe)}$, $d_{II\bot(Ga-Fe)}$ and  $d_{II\|(Ga-Fe)}$, see Table \ref{tab_Ga}).
The  \textit{y} and \textit{z} values are for all the samples less or equal to 30 $\pm$ 20. Therefore, we can say that for the samples with $x=24$, the best-fits results at the Fe K-edge are compatible with  \textit{y} and \textit{z} values expected for D0$_3$ structure.

\begin{table}
	\begin{tabular}{cc|ccc|ccc}
		&  &Ga K-edge  &  &  &Fe K-edge &   &   \\
		\hline \hline 
		\textit{Sample}    & $T_{s} $  &  $d_{1(Ga-Fe/Ga)}$ & y & $\sigma^2_{I}$ & $d_{1(Fe-Fe/Ga)}$& y & $\sigma^2_{I}$  \\
	
		&  ($^o$C)    &  (\AA)   &  (at \% Ga)    &   (\AA$^2$)    &     (\AA)  &   \textbf{} &  (\AA$^2$)    \\	\hline

		S2-400    & 400 &2.60  &0  &0.008  &2.56/2.59  &40  & 0.009  \\
		
		S2-600    & 600 &2.60  & 0 &0.007  & 2.57/2.60 &40  &  0.008  \\
		
		\hline 
		
	\end{tabular}
	\caption{Best fit results at Ga and Fe K-edge for the \textit{fcc} Fe-Ga samples.  Statistical error, calculated by the fit covariance matrix,  are equal to $0.01$~\AA\ for $d_I$,  13 for \textit{y} and 0.005 for $\sigma^2$.}
	\label{tab_fcc}
\end{table}

\subsubsection{fcc samples}

The fit approach was the same as for the \textit{bcc} samples but in this case the first FT peak is due to the contribution of a single first coordination shell of 12 NN atoms to which the fit is restricted. The lattice parameter now corresponds to the interatomic distance of the II coordination shell the contribution of which is the low intensity FT peak showing at about 3.5 \AA.  Polarized EXAFS spectra in the parallel and perpendicular to surface directions were practically identical showing that neither tetragonal deformation nor local composition anisotropy are present.   
The fitting was performed on the Fourier filtered $\chi(q)$ in the range $2-12.5 $~\AA$^{-1}$ refining
the following parameters: the origin of photoelectron energy $E_{0}$; the interatomic distances
$d_{I(Fe-Fe)}$, $d_{I(Fe-Ga)}$ (Fe K-edge) , $d_{I(Ga-Ga)}$, $d_{I(Ga-Fe)}$ (Ga K-edge), the Debye-Waller factors for I  shell and the Ga population factor shell \textit{y}. The  $E_{0}$ best fit values for the different samples were very close to each other ($E_{0}\cong$ 3 eV), showing differences lower than 0.5 eV. The  $S_{0}^{2}$  amplitude reduction factor was fixed to 0.7 for all the samples; the coordination number was kept fixed to its crystallographic value $N_{I} = 12$.

The results at the Ga and Fe K-edge are reported in Table \ref{tab_fcc}.
Regarding the Ga distribution we observe a Ga anticlustering mechanism at the Ga K-edge analogous  to that observed for the \textit{bcc} samples. No Ga atoms are found around the Ga absorbers within the fit error bar on \textit{y} ($\Delta y$=10). At the  Fe K-edge, the Ga population is found to be equal to $\approx$ 40, that is higher than the nominal Ga concentration \textit{x} = 24. It is not far from the value expected for the \textit{fcc} L1$_2$ ordered structure, compatible with these samples composition, in which  4 Ga atoms out of 12 are expected in the Fe I coordination shell (\textit{y} = 33) and no Ga atoms are foreseen in the Ga I shell.

Regarding interatomic distances we find $d_{I(Fe-Fe)}$= 2.56 - 2.57~\AA\  and $d_{I(Fe-Ga)}$ = 2.59 -2.60~\AA = $d_{I(Ga-Fe)}$.
The Fe-Fe and Fe-Ga interatomic distances show a difference of about 0.03 ~\AA, giving an average \textit{fcc} \textit{a} parameter of 3.64~\AA,  that is in fair agreement with the values found by XRD.	

\subsection{Molecular dynamics results}	


Molecular dynamics (MD) calculations ware performed,  by using the VASP code  \cite{KRESSE199615}, to obtain the equilibrium  distribution of Fe and Ga atoms in the \textit{bcc} lattice and help to interprete the EXAFS results. A (4 $\times$ 4 $\times$ 3)\textit{a}  supercell is used in the calculation containing  19 Ga atoms and 77 Fe atoms, all of them initially randomly distributed in the \textit{bcc} structure.  VASP uses DFT-based first-principles calculations with the generalized gradient approximation (GGA) for correlation and exchange \cite{PhysRevB.23.5048} and the plane basis is set on projector augmented wave (PAW) pseudopotential for describing the core electrons \cite{PhysRevB.50.17953}. The valence states of the Fe and Ga are [Ar]-3p$^{6}$3d$^{7}$4s$^{1}$ and [Ar]-3p$^{6}$3d$^{10}$4s$^{2}$4p$^{1}$, respectively, and the calculation was done at $\Gamma$ point only due to computational limitations. The MD simulation was done under canonic ensemble using the algorithm of Nos\'e-Hoover, that controls the frequency of the temperature oscillations during the simulation \cite{doi:10.1063/1.447334,PhysRevA.31.1695}. During the annealing process, up to 2000 K, the melting state was obtained at around 0.5 ps getting a stable energy of $\approx$ -645 eV/cell. After that, from this melting state, the cooling process down to 250 K was carried out using a cooling rate of 1 K/fs. Finally, the structure was optimized at 0 K by using a RMM-DIIS-Quasi-Newton method \cite{PULAY1980393} with a force on each atom $\approx$ 1-2 mRy/a.u per atom and $\approx$ 0.1 meV for the energy convergence. The final optimized cluster is analyzed to compare the Ga-Fe distribution with the results obtained by EXAFS.

MD calculations were carried out for a Ga concentration x $\approx$ 20. In order to compare the MD calculations with the EXAFS results we analyze the number of Ga-Ga pairs in first and second shell, N$_{PI}$ and N$_{PII}$ respectively in the optimized MD cluster.  If the 
Ga atoms were randomly distributed in the Fe lattice one should observe N$_{PI}$ = (8 $\times$ 0.2) $\times$ 19 = 30 and 
N$_{PII}$=(6 $\times$ 0.2) $\times$ 19 = 22.8. 
If we count the Ga-Ga pairs in the cluster  obtained by MD calculation we obtain   N$_{PI}$ = 16 and N$_{PII}$ = 10, \textit{i.e.} values that are much lower than what it can be expected for a random Ga distribution, indicating a tendency to an anticlustering of the Ga atoms.

To compare with  the \textit{y} and \textit{z} values reported in Table \ref{tab_Ga}, we normalize N$_{PI}$ and N$_{PII}$  to the total number of  pairs that Ga atoms have in the I  and II shells,  8 $\times$ 19 = 152 and 6 $\times$ 19 = 114, respectively, and multiply by 100.
Therefore, the Ga concentrations (in Ga relative units) in the I and II shell are  n$_{PI}$ = (16/152) x 100  = 10.7  and n$_{PII}$ = (10 /114) x 100 = 8.8   for the MD cluster and n$_{PI}$= (30/152) x 100 = 19.7 and n$_{PII}$ = (22.8/114)  x 100 = 20, as expected,  for a random distribution of Ga atoms. One can observe that the EXAFS values are in agreement, within the  error bars, with the concentration values obtained by the MD calculations that show the existence of a local Ga ordering in the epitaxial Fe-Ga films.

We note that for the composition studied here by MD, x $\approx$ 20, the equilibrium phase diagram, see Fig. \ref{fig:FiguraDiagramas}a, indicates a coexistence of the A2 and Ll$_2$ phases, however the calculation is performed for the  \textit{bcc} structure and the interpretation of this result can be done under the consideration of a metastable equilibrium observed in the films prepared at low temperature reported here, since the volume of the film is constrained by the substrate.

\section{Discussion}

\subsection{bcc films structure}

The EXAFS results, supported by MD calculations, show that  Ga has a clear tendency to anticlustering, \textit{i.e.} the Ga atoms tend to stay as far as possible from each other in the FeGa lattice leading to a local D0$_3$ ordering that is the kind of atomic arrangement minimizing the number of Ga-Ga pairs. 
Nevertheless, EXAFS give us a picture of the short-range-order that in our case, since we can perform a quantitative analysis of the first FT peak, is limited to a distance from the absorber equal to one single lattice parameter.
The evidence of a D0$_3$ long-range-order should be provided by XRD or TEM  but it is not the case, as reported in the previous section, since no (1/2 1/2 1/2)  reflections have been observed. On the other side, some samples (see Table \ref{tab_estruc}) present  the  (001) reflection that is a signature of chemical ordering in a \textit{bcc} structure.  The B2 structure requires a 50\% Ga content value far away from the  values obtained by EDX, see Table \ref{tab_estruc}, while the D0$_3$ structures will rise diffuse (1/2 1/2 n/2)  peaks,  not found in TEM and XRD experiments. 
These results, suggest that the films adopts a D0$_3$-like structure, which means that only  Fe atoms are present in the I and II shells, but with absence of that long-rage order.

If we compare our results with previous literature on bulk slow-cooled or quenched FeGa samples \cite{Du2010} in which a clear D0$_3$ ordered phase was observed, we can state that epitaxial growth at low temperature looks to reduce the occurrence of long-range D0$_3$ ordering. This gives to epitaxial growth a further advantage over other techniques since the formation of D0$_3$ phase is known to be detrimental for magnetostriciton  strength \cite{Srisukhumbowornchai2002}.
Also,  Pascarelli et al. \cite{Pascarelli2008}  report EXAFS results on one melt-spun FeGa sample in which about the same kind of anticlustering mechanism is observed at the Ga K-edge whereas a random distribution of Ga was found at the Fe K-edge
suggesting a less clear atomic arrangement of the Ga atoms in contrast with the samples of ref. \cite{Du2010}. 
These authors found no Ga atoms in the I coordination shell, as in our case, and 1 Ga atom out of 6 in the II one. This results was consistent with the formation of Ga-Ga pairs able to enhance MS strength. We must note that in our case all the samples are thin epitaxial layers with different growth dynamics compared with bulk samples. Our results show that for most of the samples the possibility of Ga-Ga pairs formation is less than 10\% (\textit{i.e.} the fit error bar on \textit{y}) and for one of them (S1-24a) is of the same order as in the cited paper. 
A systematic study comparing structural and MS properties in thin Fe-Ga films should be carried out to prove the role of Ga-Ga pairs often invoked but not actually verified so far.
This tendency to a local ordering mechanism is observed also for the \textit{fcc} samples and it is in agreement with the appearance of reflection (001) in complementary XRD experiments reported in previous section. 
It is consistent with the short-range-ordering mechanism observed in the \textit{bcc} samples obeying to an anticlustering tendency of the Ga atoms in the FeGa lattice. 
Our results also show that the driving force of the ordering mechanism is not the residual mismatch strain, since samples with different strain content show the same local Ga ordering.

\subsection{fcc films growth}

The experimental data reported here show that \textit{fcc} L1$_2$ films with the (001) growing plane can be obtained directly onto the MgO(001) surface without the need of time consuming  heat treatments, despite of the larger misfit between MgO and th
e \textit{fcc} structure. However, in other systems, with large lattice mismatch between the lattice parameters of over-layer and substrate, epitaxial films has been obtained, for instance Ni(100) \textit{fcc} films onto  MgO(100) (misfit  $\sim$ 15\%) if the Ni is deposited with the substrate heated from 100 $^o$C to 200 $^o$C \cite{Ni_MgO}  or  Nb  on Al$_2$O$_3$\cite{0305-4608-12-6-001}. 
Instead of being due to the matching  condition
with the substrate, 
this heteroepitaxy is partially explained as the result of an
adjustment between surpercells involving a different number of unit cells for film and substrate. Thus,  here for the S2-600 film with \textit{a} = 3.697 \AA, a cell with 8 atomic distances (1.847 \AA\ $\times$ 8 = 14.79 \AA) facing a MgO cell with seven atomic distances (2.107 \AA\ $\times$ 7 = 14.75 \AA ) will have a small effective mismatch of $\sim$ 0.3 \%.

\section{Conclusions}

We have shown that the Fe-Ga films grown at low (150 $^o$ 
C) temperature present \textit{bcc} structure with anticlustering ordering for the Ga atoms and chemical superorder for compositions above \textit{x} $\approx$ 17, without the presence of long range D0$_3$ order. The equilibrium Ll$_2$ crystal phase is obtained for a composition around \textit{x} = 25 for T$_s$ larger that 400 $^o$C. The information is summarized in a metastable phase diagram composition \textit{vs.} growing temperature.

\section{Acknowledgments}
This work has been supported by Spanish MICINN (Grants No. MAT2012-31309 and MAT2015-66726-R), MECD (Programa Campus de Excelencia Internacional Iberus) and Gobierno de Arag\'{o}n (Grant E10-17D) and Fondo Social Europeo. Authors would like to acknowledge the use of Servicio General de Apoyo a la Investigación-SAI, Universidad de Zaragoza. We acknowledge the use of the microscopy infrastructure available in the Laboratorio de Microscopías Avanzadas (LMA) at Instituto de Nanociencia de Arag\'{o}n (University of Zaragoza, Spain). We also thanks Dr. Antonio Bad\'{\i}a at Universidad de Zaragoza the use of the SIE cluster in MD calculations. The XANES experiments were performed on beamline BM30D at the European Synchrotron Radiation Facility (ESRF), Grenoble, France and on beamline XAFS at Elettra, Trieste, Italy. We are grateful to Isabelle  Kieffer at the ESRF (beamline BM30B)  and,  Giuliana Aquilanti  and Clara Guglieri at Elettra (beamline XAFS) for providing assistance in using synchrotron facilities. AB thanks MINECO for the Ph. D. grant BES-2016-076482.
\bibliographystyle{unsrt}



\section{Bibliography}
\begin{thebibliography}{10}
	
	\bibitem{Clark2000}
	Arthur~E. Clark, James~B. Restorff, Marilyn Wun-Fogle, Thomas~A. Lograsso, and
	Deborah~L. Schlagel.
	\newblock Magnetostrictive properties of body-centered cubic {Fe-Ga} and
	{Fe-Ga-Al} alloys.
	\newblock {\em IEEE Trans Magn}, 36(5 I):3238--3240, 2000.
	
	\bibitem{Clark2001}
	A.~E. Clark, M.~Wun-Fogle, J.B. Restorff, T.A. Lograsso, and J.R. Cullen.
	\newblock Effect of quenching on the magnetostriction on
	{Fe}$_{1-x}${Ga}$_{x}$.
	\newblock {\em IEEE Trans Magn}, 37(4):2678--2680, 2001.
	
	\bibitem{Clark2003}
	A.~E. Clark, K.~B. Hathaway, M.~Wun-Fogle, J.~B. Restorff, T.~A. Lograsso,
	V.~M. Keppens, G.~Petculescu, and R.~A. Taylor.
	\newblock Extraordinary magnetoelasticity and lattice softening in bcc {Fe-Ga}
	alloys.
	\newblock {\em J Appl Phys}, 93(10 3):8621--8623, 2003.
	
	\bibitem{Atulasimha2011}
	Jayasimha Atulasimha and Alison~B Flatau.
	\newblock A review of magnetostrictive iron-gallium alloys.
	\newblock {\em Smart Mater. Struct.}, 20(4):043001, 2011.
	
	\bibitem{NatureFeGa}
	Harsh~Deep Chopra and Manfred Wuttig.
	\newblock Non-{J}oulian magnetostriction.
	\newblock {\em Nature}, 521(7552):340--343, 2015.
	
	\bibitem{Parkes2013}
	D~E Parkes, L~R Shelford, P~Wadley, V~Hol\'{y}, M~Wang, A~T Hindmarch,
	G~van~der Laan, R~P Campion, K~W Edmonds, S~A Cavill, and A~W Rushforth.
	\newblock {Magnetostrictive thin films for microwave spintronics.}
	\newblock {\em Sci Rep}, 3:2220, 2013.
	
	\bibitem{Onuta2011}
	Tiberiu~Dan Onuta, Yi~Wang, Christian~J. Long, and Ichiro Takeuchi.
	\newblock Energy harvesting properties of all-thin-film multiferroic
	cantilevers.
	\newblock {\em Appl Phys Lett}, 99(20):10--13, 2011.
	
	\bibitem{Hu2011}
	Jia-Mian Hu, Zheng Li, Long-Qing Chen, and Ce-Wen Nan.
	\newblock High-density magnetoresistive random access memory operating at
	ultralow voltage at room temperature.
	\newblock {\em Nat. Commun.}, 2:553, 2011.
	
	\bibitem{PhysRevLett.104.197201}
	Sen Yang, Huixin Bao, Chao Zhou, Yu~Wang, Xiaobing Ren, Yoshitaka Matsushita,
	Yoshio Katsuya, Masahiko Tanaka, Keisuke Kobayashi, Xiaoping Song, and
	Jianrong Gao.
	\newblock Large magnetostriction from morphotropic phase boundary in
	ferromagnets.
	\newblock {\em Phys. Rev. Lett.}, 104:197201, May 2010.
	
	\bibitem{PhysRevLett.111.017203}
	Richard Bergstrom, Manfred Wuttig, James Cullen, Peter Zavalij, Robert Briber,
	Cindi Dennis, V.~Ovidiu Garlea, and Mark Laver.
	\newblock Morphotropic phase boundaries in ferromagnets:
	{Tb}$_{1-x}${Dy}$_{x}${Fe}$_{2}$ alloys.
	\newblock {\em Phys. Rev. Lett.}, 111:017203, Jul 2013.
	
	\bibitem{Ma2017}
	Tianyu Ma, Junming Gou, Shanshan Hu, Xiaolian Liu, Chen Wu, Shuai Ren, Hui
	Zhao, Andong Xiao, Chengbao Jiang, Xiaobing Ren, and Mi~Yan.
	\newblock Highly thermal-stable ferromagnetism by a natural composite.
	\newblock {\em Nat. Commun.}, 8:13937--, January 2017.
	
	\bibitem{Handbook}
	Gabriela Petculescu, Ruqian Wu, and Robert McQueeney.
	\newblock {\em Magnetoelasticity of bcc {Fe-Ga} alloys}, volume~20, chapter~3.
	\newblock Elsevier, 2012.
	
	\bibitem{inCollFeGa}
	H.~Okamoto.
	\newblock {FeGa}.
	\newblock In {\em Binary Alloy phase diagrams}, page 1702. Materials
	Information Society, second edition, 1992.
	
	\bibitem{Srisukhumbowornchai2002}
	N.~Srisukhumbowornchai and S.~Guruswamy.
	\newblock {Influence of ordering on the magnetostriction of {Fe}-27.5 at. %
		{Ga} alloys}.
	\newblock {\em J. Appl. Phys.}, 92(9):5371--5379, 2002.
	
	\bibitem{IKEDA2002198}
	O~Ikeda, R~Kainuma, I~Ohnuma, K~Fukamichi, and K~Ishida.
	\newblock Phase equilibria and stability of ordered b.c.c. phases in the
	{Fe}-rich portion of the {Fe–Ga} system.
	\newblock {\em J. Alloys Compd.}, 347(1):198 -- 205, 2002.
	
	\bibitem{Liu1}
	C.~Liu, E.~R. Moog, and S.~D. Bader.
	\newblock Polar kerr-effect observation of perpendicular surface anisotropy for
	ultrathin fcc {Fe} grown on {Cu}(100).
	\newblock {\em Phys. Rev. Lett.}, 60:2422--2425, Jun 1988.
	
	\bibitem{McClure2009}
	Adam McClure, S.~Albert, T.~Jaeger, H.~Li, P.~Rugheimer, J.~a. Schaefer, and
	Y.~U. Idzerda.
	\newblock Properties of single crystal {Fe}$_{1-x}${Ga}$_{x}$ thin films.
	\newblock {\em J. Appl. Phys.}, 105(7):07A938, 2009.
	
	\bibitem{Eddrief2011}
	M.~Eddrief, Y.~Zheng, S.~Hidki, B.~{Rache Salles}, J.~Milano, V.~H. Etgens, and
	M.~Marangolo.
	\newblock Metastable tetragonal structure of {Fe}$_{100-x}${Ga}$_x$ epitaxial
	thin films on {ZnSe/GaAs}(001) substrate.
	\newblock {\em Phys. Rev. B}, 84(16):1--5, 2011.
	
	\bibitem{Beardsley2017}
	R.~P. Beardsley, D.~E. Parkes, J.~Zemen, S.~Bowe, K.~W. Edmonds, C.~Reardon,
	F.~Maccherozzi, I.~Isakov, P.~A. Warburton, R.~P. Campion, B.~L. Gallagher,
	S.~A. Cavill, and A.~W. Rushforth.
	\newblock Effect of lithographically-induced strain relaxation on the magnetic
	domain configuration in microfabricated epitaxially grown
	{F}e$_{81}${G}a$_{19}$.
	\newblock {\em Sci. Rep.}, 7:42107, February 2017.
	
	\bibitem{Oka2001}
	H~Oka, A~Subagyo, M~Sawamura, K~Sueoka, and K~Mukasa.
	\newblock Scanning tunneling spectroscopy of c(2 x 2) reconstructed {Fe}
	thin-film surfaces.
	\newblock {\em Jpn. J. Appl. Phy.}, 40(6B):4334--4336, 2001.
	
	\bibitem{Luo2011}
	Yuansu Luo, Philipp Leicht, Aleksej Laptev, Mikhail Fonin, Ulrich R{\"u}diger,
	Markus Laufenberg, and Konrad Samwer.
	\newblock Effects of film thickness and composition on the structure and
	martensitic transition of epitaxial off-stoichiometric {Ni-Mn-Ga} magnetic
	shape memory films.
	\newblock {\em New J. Phys.}, 13(1):013042, 2011.
	
	\bibitem{Kawamiya}
	Nobuo Kawamiya, Kengo Adachi, and Yoji Nakamura.
	\newblock Magnetic properties and m\"{o}ssabauer investigations of {Fe-Ga}
	alloys.
	\newblock {\em Journal of the Physical Society of Japan}, 33(5):1318--1327,
	1972.
	
	\bibitem{He2016177}
	Yangkun He, Chengbao Jiang, Wei Wu, Bin Wang, Huiping Duan, Hui Wang, Tianli
	Zhang, Jingmin Wang, Jinghua Liu, Zaoli Zhang, Plamen Stamenov, J.M.D. Coey,
	and Huibin Xu.
	\newblock Giant heterogeneous magnetostriction in {Fe–Ga} alloys: Effect of
	trace element doping.
	\newblock {\em Acta Mater.}, 109:177 -- 186, 2016.
	
	\bibitem{DoublePeak}
	D.~Van Dyck, G.~Van Tendeloo, and S.~Amelinckx.
	\newblock Diffraction effects due to a single translation interface in a small
	crystal.
	\newblock {\em Ultramicroscopy}, 15:357--370, 1984.
	
	\bibitem{Rehr}
	J. J. Rehr and R. C. Albers. 
	\newblock Theoretical approaches to x-ray absorption fine structure
	\newblock {\em Rev. Mod. Phys}, 72(3):621--654, 2000.
	
	\bibitem{Boscherini-Chap7-Lamberti-Agostini-2012}
	F.~Boscherini.
	\newblock In G.~Agostini and C.~(Eds) Lamberti, editors, {\em Characterization
		of Semiconductor Heterostructures and Nanostructures II}, Chapter~9. Amsterdam, 2012.
	Elsevier.
	
	\bibitem{brouder_ang-dep_1990}
	C~Brouder.
	\newblock Angular dependence of {X}-ray absorption spectra.
	\newblock {\em J. Phys.: Condens. Matter}, 2(3):701, 1990.
	
	\bibitem{Ravel2}
	B.~Ravel.
	\newblock {ATOMS}: crystallography for the {X}-ray absorption spectroscopist.
	\newblock {\em J. Synchrotron Rad.}, 8:314--316, 2001.
	
	\bibitem{ankudinov_real-space_1998}
	A.~L. Ankudinov, B.~Ravel, J.~J. Rehr, and S.~D. Conradson.
	\newblock Real-space multiple-scattering calculation and interpretation of
	{X}-ray-absorption near-edge structure.
	\newblock {\em Phys. Rev. B}, 58(12):7565--7576, September 1998.
	
	\bibitem{ravel_athena_2005}
	B.~Ravel and M.~Newville.
	\newblock {ATHENA}, {ARTEMIS}, {HEPHAESTUS}: data analysis for {X}-ray
	absorption spectroscopy using {IFEFFIT}.
	\newblock {\em J. Synchrotron Rad.}, 12(4):537--541, June 2005.
	
	\bibitem{newville_ifeffit_2001}
	Matthew Newville.
	\newblock {IFEFFIT}: interactive {XAFS} analysis and {FEFF} fitting.
	\newblock {\em J. Synchrotron Rad.}, 8(2):322--324, March 2001.
	
	\bibitem{Pascarelli2008}
	S.~Pascarelli, M.~Ruffoni, R.~{Sato Turtelli}, F.~Kubel, and R.~Gr\"{o}ssinger.
	\newblock {Local structure in magnetostrictive melt-spun {Fe}$_{80}${Ga}$_{20}$
		alloys}.
	\newblock {\em Phys. Rev. B}, 77(18):1--8, 2008.
	
	\bibitem{Du2010}
	Y.~Du, M.~Huang, S.~Chang, D.~L. Schlagel, T.~A. Lograsso, and R.~J. McQueeney.
	\newblock Relation between {Ga} ordering and magnetostriction of {Fe-Ga} alloys
	studied by {X}-ray diffuse scattering.
	\newblock {\em Phys Rev B}, 81(5):1--9, 2010.
	
	\bibitem{KRESSE199615}
	G.~Kresse and J.~Furthm\"uller.
	\newblock Efficiency of ab-initio total energy calculations for metals and
	semiconductors using a plane-wave basis set.
	\newblock {\em Comput. Mater. Sci.}, 6(1):15 -- 50, 1996.
	
	\bibitem{PhysRevB.23.5048}
	J.~P. Perdew and Alex Zunger.
	\newblock Self-interaction correction to density-functional approximations for
	many-electron systems.
	\newblock {\em Phys Rev B}, 23:5048--5079, May 1981.
	
	\bibitem{PhysRevB.50.17953}
	P.~E. Bl\"{o}chl.
	\newblock Projector augmented-wave method.
	\newblock {\em Phys. Rev. B}, 50:17953--17979, Dec 1994.
	
	\bibitem{doi:10.1063/1.447334}
	Shuichi Nos\'e.
	\newblock A unified formulation of the constant temperature molecular dynamics
	methods.
	\newblock {\em J. Chem. Phys.}, 81(1):511--519, 1984.
	
	\bibitem{PhysRevA.31.1695}
	William~G. Hoover.
	\newblock Canonical dynamics: Equilibrium phase-space distributions.
	\newblock {\em Phys. Rev. A}, 31:1695--1697, Mar 1985.
	
	\bibitem{PULAY1980393}
	P.~Pulay.
	\newblock Convergence acceleration of iterative sequences. the case of scf
	iteration.
	\newblock {\em Chem. Phys. Lett.}, 73(2):393 -- 398, 1980.
	
	\bibitem{Ni_MgO}
	E.B. Svedberg, P.~Sandstr\"om, J.-E. Sundgren, J.~E. Greene, and L.D. Madsen.
	\newblock Epitaxial growth of {Ni on MgO(002)} : surface interaction vs.
	multidomain strain relief.
	\newblock {\em Surf Sci}, 429:206--216, 1999.
	
	\bibitem{0305-4608-12-6-001}
	S~M Durbin, J~E Cunningham, and C~P Flynn.
	\newblock Growth of single-crystal metal superlattices in chosen orientations.
	\newblock {\em J. Phys. F: Met. Phys.}, 12(6):L75, 1982.
	
\end{thebibliography}

\end{document}